\documentclass[11pt]{article}
\usepackage{hyperref}      
\pdfoutput=1
\begin{document}

\title{Initial Stages of Wind-Waves Evolution, \\ Temporal vs. Spatial Cases}
\author{\centerline {Andrey Zavadsky$^{\textrm{1}}$, Dan Liberzon$^{\textrm{2}}$ Lev Shemer$^{\textrm{1}}$}\\
\\\vspace{6pt} \centerline{$^{\textrm{1}}$School of Mechanical Engineering, Tel Aviv University,} \\ Tel Aviv 69978, Israel \\ $^{\textrm{2}}$Civil Engineering and Geological Sciences, University of Notre Dame, \\ Notre Dame, IN, 46556, USA}
\maketitle

\begin{abstract}
  The video describes initial stages spatial and temporal evolution of wind generated waves. This fluid dynamics video was created at Tel Aviv University small scale wind-wave flume as a part of an ongoing experimental program aimed at gaining better understanding of complex processes governing the excitation of water waves and their evolution in the presence of wind. Despite many decades of intense research mechanisms governing water waves’ generation by wind and their evolution in space and time are still not fully understood. 
\end{abstract}

\section{Introduction} 
The small-scale wind-wave flume consisting of a closed loop wind tunnel over a wave flume allows generation of strong turbulent wind at velocities comparable with those measured in the open sea. A centrifugal, 7.5 HP, blower drives the air flow attaining mean air flow velocities in the test section of up to 15 m/s. Maximum mean wind velocity in the present video was 12 m/s. The 5 m long, 0.4 m wide, and 0.5 m deep test section is made of clear, reinforced, 6 mm thick glass plates. Clear glass walls of the tank mounted on the frame allow viewing of the flow from any desired angle. The test section is sufficiently long to conduct studies of initial stages of spatial water-waves evolution and enables an easy access to the test section for instrumentation installation. Rectangular air inlet and outlet openings in the tank are 0.4 m wide and 0.25 m high, the water depth of about 0.2 m in the wave channel satisfies the deep water conditions. A 0.4 m long flap ensures smooth air flow from the settling chamber into the test section. The relatively small size of the experimental facility allows accurate control of experimental conditions. For more details about the experimental facility see Liberzon and Shemer (2011). 

\section{Video Description}
The first part of the video, titles "Temporal Evolution", demonstrates record of the initially calm water surface variation from the moment the wind tunnel blower is switched on. The wind velocity measured by a Pitot tube and the instantaneous surface elevation determined by a capacitance-type wave gauge are recorded simultaneously with the imaging of the wave field. The video camera in this part is located at a fixed position at some angle to the wind-wave flume; the field of view spans for about 2 m starting from the flap. The Pitot tube record demonstrates that it takes about 4 s until a quasi-steady wind velocity is attained in the test section, first visible ripples appear roughly at the same time. For few seconds the wave amplitudes grow in time but the wave field remains independent of fetch and statistically spatially homogeneous. These finding are in general agreement with Caulliez et al. (1998). Larson and Wright (1974) reported on initial temporal growth rates following impulsively applied wind forcing for short waves using radar backscatter cross-section measurements. Their experiments were carried out in a facility similar in size to the present wind-wave flume. One can therefore expect that the typical time scales were also similar. The present video demonstrates, however, that reliable data on purely temporal growth of the wave field can only be accumulated during a quite short period comparable to the response time of the system. The wave gauge output demonstrates that after about 12 s following the operation of the blower the wave field attains a quasi-steady state and becomes strongly fetch-dependent. It is therefore advantageous to study spatial rather than temporal evolution of the wind-wave field. Numerous studies of spatial growth rates were carried out, for a recent review see Liberzon and Shemer (2011). \\ In the second part of the video, called "Spacial Evolution", the camera moves along the test section and follows waves propagation along the flume allowing examination of the initial stages of the spatial evolution of wind waves as they grow in amplitude and length. The recording is performed after the steady state has been achieved. The main frame shows the air water interface as it varies with fetch along the test section. The second, smaller, frame in the bottom left corner illustrates the process of the interface capturing as the camera and the illumination equipment move on a rail along the test section at a constant speed. The imaged area in the main frame is about 0.25 m long. The wave field in the proximity to the entrance to the text section is dominated by short ripples that are couple of cm long. As the camera moves along the flume, the waves gradually become longer and higher. Towards the far end of the test section the characteristic wave lengths become comparable with that of the imaged window. Waves become steeper and some breaking events can be observed. At the end of the test section waves energy is damped at the beach. 

\section{References}
1. Caulliez, G., Ricci N. and Dupont, R. 1998 The generation of the first visible wind waves. \textit{Phys. Fluids} \textbf{10}, 4, 757-759. \\ 2. Larson, T. R. and Wright, J. W. 1974 Wind-generated gravity capillary waves: laboratory measurements of temporal growth rates using microwave backscatter. \textit{J. Fluid Mech.} \textbf{70}, 417–436. \\ 3. D. Liberzon and L. Shemer (2011) Experimental study of the initial stages of wind waves' spatial evolution. \textit{J. Fluid Mech.} \textbf{681}, 462-498, doi:10.1017/jfm.2011.208.

\end{document}